\documentclass[multphys,vecphys]{svmult}

\usepackage{makeidx}         
\usepackage{graphicx}        
\usepackage{multicol}        
\usepackage[bottom]{footmisc}

\makeindex             

\newcommand{\be}{\begin{equation}}
\newcommand{\ee}{\end{equation}}
\newcommand{\bn}{\begin{eqnarray}}
\newcommand{\en}{\end{eqnarray}}
\newcommand{\bd}{\begin{displaymath}}
\newcommand{\ed}{\end{displaymath}}
\newcommand{\bnn}{\begin{eqnarray*}}
\newcommand{\enn}{\end{eqnarray*}}
\def\Ref#1{(\ref{#1})}
\def\Journal#1#2#3#4#5{#1: \ #2 \ {\bf #3}, \  #4 \ (#5)}
\def\journal#1#2#3#4#5{#1: \ #2 \ {\bf #3}, \ #4 \ (#5)}

\begin{document}
\title*{Manifestation of Chaos in Real Complex Systems: Case of Parkinson's Disease}
\titlerunning{Manifestation of Chaos in Real Complex Systems}
\author{Renat M. Yulmetyev\inst{1}\and
Sergey A. Demin\inst{1}\and Peter H\"{a}nggi\inst{2}}
\institute{Department of Physics, Kazan State Pedagogical
University, 420021 Kazan, Mezhlauk Street, 1 Russia
\texttt{rmy@theory.kazan-spu.ru} \and Department of Physics,
University of Augsburg, Universit\"atsstrasse 1, D-86135 Augsburg,
Germany}
%
%
\maketitle In this chapter we present a new approach to the study
of manifestations of chaos in real complex system. Recently we
have achieved the following result. In real complex systems the
informational measure of chaotic chatacter (IMC) can serve as a
reliable quantitative estimation of the state of a complex system
and help to estimate the deviation of this state from its normal
condition. As the IMC we suggest the statistical spectrum of the
non-Markovity parameter (NMP) and its frequency behavior. Our
preliminary studies of real complex systems in cardiology,
neurophysiology and seismology have shown that the NMP has diverse
frequency dependence. It testifies to the competition between
Markovian and non-Markovian, random and regular processes and
makes a crossover from one relaxation scenario to the other
possible. On this basis we can formulate the new concept in the
study of the manifestation of chaoticity. We suggest the
statistical theory of discrete non-Markov stochastic processes to
calculate the NMP and the quantitative evaluation of the IMC in
real complex systems. With the help of the IMC we have found out
the evident manifestation of chaosity in a normal (healthy) state
of the studied system, its sharp reduction in the period of
crises, catastrophes and various human diseases. It means that one
can appreciably improve the state of a patient (of any system) by
increasing the IMC of the studied live system. The given
observation creates a reliable basis for predicting crises and
catastrophes, as well as for diagnosing and treating various human
diseases, Parkinson's disease in particular.

\section{Introduction}
Today the study of manifestations of chaos in real complex systems
of diverse nature  has acquired great importance. The analysis of
some properties and characteristics of real complex systems is
impossible without quantitative estimate of various manifestation
of chaos. The dynamics or evolution of the system can be predicted
by the change of its chaoticity or regularity. The discovery of
the phenomenon of chaos in dynamic systems has changed the
attitude with regard to the functioning of complex systems, a
human organism in particular. The chaos is the absence of
regularity. It characterizes the randomness and the
unpredictability of the changes of the behavior of a system. At
the same time, the presence of chaos in dynamic systems does not
mean it cannot be taken under control. Instability of dynamic
systems in the state of chaos creates special sensitivity to both
external and internal influences and perturbations. The series of
weak perturbations of the parameters of the system allows to
change its characteristics in the required direction. "Chaos" is
frequently understood as a determined dynamic chaos, that is, the
dynamics dependent on the initial conditions, parameters.

Lasers, liquid near to a threshold of turbulence, devices of
nonlinear optics, chemical reactions, accelerators of particles,
classical multipartite systems, some biological dynamic models are
the examples of nonlinear systems with the manifestation of
determined chaos. Now manifestations of chaos are being studied in
different spheres of human activity.

The control of the behavior of chaotic systems is one of the most
important problems. Most of the authors see two basic approaches
to solve the problems \cite {Boc, Touc}. Both directions envisage
a preliminary choice of a certain perturbation. The selected
perturbation is used to exert influence on the chaotic system. The
first direction relies on an internal perturbation, the choice of
which is based on the state of the system. The perturbation
changes the parameter or the set of parameters of the system,
which results in the ordered behaviour of the chaotic systems. The
methods focusing on the choice of such parameters (perturbations)
are referred to as "methods with a feedback" \cite{Boc}-\cite
{Plapp}. They do not dependent on the studied chaotic system
(model) as these parameters can be selected by observing the
system for some period of time. One also considers that the
methods with a delayed feedback \cite {Pyr, Just} belong to the
first direction. The second approach presupposes that the choice
of the external perturbation does not dependent on the state of
the chaotic system under consideration. By affecting the studied
system with the similar perturbation, it is possible to change its
behaviour. The present group of methods is an alternative to the
first one. These methods can be used in cases when internal
parameters depend on the environment \cite {Boc, Lima, Braim}.

Generally, when choosing internal (external) perturbations it is
possible to determine three basic stapes: the estimation of the
initial information, the choice of the perturbation and the
bringing the chosen strategy of control into action (its practical
realization). At the first stage  the information on the state of
the studied system is collected. At the second stage the received
information is processed according to the plan or strategy of the
control. On the basis of the achieved results the decision on the
choice of the internal (of the external) perturbation is accepted.
After that the chosen strategy of chaos control is put into
practice \cite {Touc}.

The initial idea of the present concept was to separate Markov
(with short-range time memory) and non-Markov (with long-range
time memory) stochastic processes. However, the study of real
complex systems has revealed additional possibilities of the given
parameter. Actually, the non-Markovity parameter represents a
quantitative measure of chaoticity or regularity of various states
of the studied system. An increase of the given parameter ($
\varepsilon_1 (0) \gg 1 $) corresponds to an increase of the
chaoticity of the state of the system. A decrease of the
non-Markovity parameter means a greater ordering (regularity) of
the state of the system. The given observation allows one to
define a new strategy for estimating of chaoticity in real
systems. This new approach in chaos theory can be presented as an
alternative to the existing methods. Further analysis of the
non-Markovity parameter allows one to define the degree of
chaoticity or regularity of a state of the system.

In this work the new strategy for  the study of manifestations of
chaoticity is applied to real complex systems. The possibilities
of the new approach are revealed at the analysis of the
experimental data on various states of a human organism with
Parkinson's disease. Parkinson's disease is a chronic progressing
disease of the brain observed in 1-2 \% of elderly people. The
given disease was described in 1817 by James Parkinson in the book
"An essay on the shaking palsy". In 19th century the French
neurologist Pierre Marie Charcot  called this disease "Parkinson's
disease". The steady progress of the symptoms and yearly
impairment of motor function is typical of Parkinson's disease.
Complex biochemical processes characteristic of Parkinson's
disease result in a lack of dopamine, a chemical substance which
is carrier signals from one nerve cell to another. The basic
symptoms typical of Parkinson's disease form the so-called
classical triad: tremor, rigidity of muscles (disorder of speech,
amimia), and depression (anxiety, irritability, apathy). The
disease steadily progresses and eventually the patient becomes a
helpless invalid. The existing therapy comprises a set of three
basic treatments: medical treatment, surgical treatment and
electromagnetic stimulation of the affected area of the brain with
the help of an electromagnetic stimulator. Today this disease is
considered practically incurable. The treatment of patients with
Parkinson's disease requires an exact estimate of the current
state of the person. The offered concept of research of
manifestations of chaoticity allows one to track down the least
changes in the patient with the help of an exact quantitative
level of description.

Earlier we found out an opportunity for defining the
predisposition of a person to the frustration of the central
nervous system due to Parkinson's disease \cite {Yulm4}. Our work
is an expansion and development of the informational possibilities
of the statistical theory of discrete non-Markov random processes
and the search for parameters affecting the health of a subject.

\section{The statistical theory of discrete non-Markov random processes.
Non-Markovity parameter and its frequency spectrum}

The statistical theory of discrete non-Markov random processes
\cite{Yulm1}-\cite{Yulm3} forms a mathematical basis for our study
of complex live systems. The theory allows one to calculate the
wide quantitative set of dynamic variables, correlation functions
and memory functions, power spectra, statistical non-Markovity
parameter, kinetic and relaxation parameters. The full
interconnected set of these variables, functions and parameters
creates a quantitative measure of chaoticity used for the
description of processes, connected with functioning of alive
organism.

We use the non-Markovity parameter  $ \varepsilon $ as a
quantitative estimate of  the non-Markov properties of the
statistical system. The non-Markovity parameter allows to
statistical processes into Markov processes ($
\varepsilon\rightarrow\infty $), quasi-Markov processes ($
\varepsilon  > 1 $) and non-Markov processes ($\varepsilon\sim$1).
Besides the non-Markovity parameter we also use the spectrum of
the non-Markovity parameter. We define the spectrum as the set of
all values of the physical parameter used for describing the state
of a system or a process. Let's consider the first and the $n$th
kinetic equations of the chain of connected non-Markov
finite-difference kinetic equations \cite{Yulm1, Yulm2}: \bn \frac
{\Delta a (t)} {\Delta t} = \lambda_1 a (t)-\tau \Lambda_1 \sum _
{j=0} ^ {m-1} M_1 (j\tau) a (t-j\tau),
\\ \cdots \nonumber \\\frac {\Delta M_n (t)} {\Delta t} = \lambda _ {n+1}
M_n (t) -\tau \Lambda _ {n+1} \sum _ {j=0} ^ {m-1} M _ {n+1}
(j\tau) M_n (t-j\tau).\nonumber\label {f1} \en The first equation
is based on the Zwanzig'-Mori's kinetic equation in nonequilibrium
statistical physics: \be \frac {d a (t)} {d t} =
-\Omega^{2}_{1}\int_ {0} ^ {t} d\tau M_1 (j\tau) a (t-j\tau)
\nonumber.\ee Here $a (t) $ is a normalized time correlation
function (TCF): \be \lim_{t \to 0} a(t)=1, \lim_{t \to \infty}
a(t)=0.\nonumber\ee The zero memory function $a(t) $ and the first
order memory function $M_1(t)$ in (1): \bn M_0 (t) =a (t) = \frac
{< {\bf A} _k^0 (0) {\bf A}_{m+k} ^m (t) >} {< | {\bf A} _k^0 (0)
| ^2 >}, ~~ t=m\tau, \nonumber \\ M_1(j \tau)= \frac{< {\bf A}_k^0
(0) \hat L_{12} \{ 1+i\tau \hat L _ {22} \} ^j \hat L _ {21} {\bf
A} _k^0 (0) >} {< {\bf A}_k^0 (0) \hat L _ {12}
\hat L _ {21} {\bf A} _k^0 (0) >}, ~~ M_1 (0) =1. \nonumber \\
{\bf A}_k^0 (0) =(\delta x_0,
\delta x_1, \delta x_2, \cdots, \delta x _ {k-1}), \nonumber \\
{\bf A}_ {m+k} ^m (t) = \{ \delta x_m , \delta x _ {m+1}, \delta
x_{m+2}, \cdots, \delta x _ {m+k-1} \}, \nonumber \en describes
statistical memory in complex systems with a discrete time (${\bf
A}_k^0 (0)$ and ${\bf A}_ {m+k} ^m (t)$ are the vectors of the
initial and final states of the studied system). The operator
$\hat L$ is a finite-difference operator: \bn i \hat L =
\frac{\bigtriangleup}{\bigtriangleup t}, \ \ \bigtriangleup t =
\tau, \nonumber \en where $\tau$ is a discretization time  step, $
\hat L_{ij}= \Pi_{i} \hat L \Pi_{j}$ ($i, j=1,2$) are matrix
elements of the splittable Liouville's quasioperator,  $
\Pi_{1}=\Pi, \Pi_{2}= P = 1-\Pi$ and $\Pi$ are projection
operators.

Let's define the relaxation times of the initial TCF and of the
first-order memory functions as follows $M_1(t)$: \be \tau_a = Re
\int_{0}^{\infty} a(t) dt, \tau_{M_1} = Re \int_{0}^{\infty}
M_1(t) dt, \cdots, \tau_{M_n} = Re \int_{0}^{\infty} M_n(t)
dt.\nonumber\ee

Then the spectrum of the non-Markovity parameter  $\{
\varepsilon\}$ is defined as an infinite set of dimensionless
numbers: \bn
\{\varepsilon_i\}=\{\varepsilon_1,\varepsilon_2,...,\varepsilon_n,...\},\nonumber
\\ \varepsilon_1=\tau_a/\tau_{M_1},
\varepsilon_2=\tau_{M_1}/\tau_{M_2}, \cdots,
\varepsilon_n=\tau_{M_n}/\tau_{M_{n+1}},\nonumber
\\ \varepsilon=\tau_{rel}/\tau_{mem}.\label {f2}\en Note that
$a(t)=M_0(t).$ The number $ \varepsilon_n $ characterizes the
ratio of the relaxation times of the memory functions $M_n$ and $
M _ {n+1} $. If for some $n $ the value of the parameter $
\varepsilon_n\to\infty $, then this  relaxation level  is
Markovian. If $ \varepsilon_n $ changes in limits from zero to a
unit value, then the relaxation level is defined as non-Markovian.
The times $ \tau _ {rel} $ (relaxation time) and $ \tau_{mem} $
(memory life time) appear when the effects of the statistical
memory in the complex discrete system are taken into account by
means of the Zwanzig'-Mori's method of kinetic equations. Thus,
the non-Markovity parameter spectrum is defined by the stochastic
properties of the TCF.

In \cite {Yulm1} the concept of generalized non-Markovity
parameter  for a frequency - dependent case was introduced: \be
\varepsilon_i (\omega) = \left \{ \frac {\mu _ {i-1} (\omega)}{
\mu_i (\omega)} \right\}^{\frac {1} {2}}. \label{f3} \ee Here as
$\mu_i(\omega)$ we have the frequency power spectrum of the i$th$
memory functions: $\mu_1(\omega)=|Re\int_{0}^{\infty}
M_1(t)e^{i\omega t}dt|^2, \cdots,
\mu_i(\omega)=|Re\int_{0}^{\infty} M_i(t) \\ e^{i\omega t}dt|^2
\nonumber.$

The use of  $ \varepsilon_i (\omega) $ allows one to find the
details of the frequency behaviour of the power spectra of the
time correlations and memory functions.

\section{The universal property of  informational manifestation of
chaoticity in complex systems}

In our work the discussion of manifestation of chaoticity is
carried out on the basis of a statistical invariant which includes
a quantitative informational measure of chaoticity and pathology
in a covariant form. The existence of this invariant is very
important for the taking decisions in the problems related to
medicine as well as for analysing a wide area of physical problems
related to complex systems of various nature.

In each live organism there is a universal informational property
of the following form: \be IMC + IMP = Invariant. \label {f4} \ee
Here IMC is an informational (quantitative) measure of chaoticity
for the concrete live system, IMP is an informational measure of a
pathological state of a live organism. As an informational
(quantitative) measure of the degree of chaoticity (regularity) we
propose to use the first point of the non-Markovity parameter  at
zero frequency: $\varepsilon_1 (0) = \left\{\frac {\mu_0 (0)}
{\mu_1(0)} \right\}^{\frac{1}{2}} $. The physical sense of the
parameter consists in comparison the relaxation scales of the time
correlation function ($a (\omega) $) to the memory functions of
the first order ($ \mu (\omega) $). Depending on the values of
this parameter one can discriminate Markov processes  (with
short-range memory) and non-Markov processes (with effects
long-range memory). Thus, the phenomena distinguished by the
greatest chaoticity correspond to Markov processes. Non-Markov
processes are connected with greater regularity. The informational
measure of a pathological state (IMP) defines the qualitative
state of a real live system.

The quantitative estimate IMC of the degree of the chaoticity of
system contains the information on a pathological state of the
system. It testifies to the close interrelation of the given
quantities. A high degree of chaoticity is characteristic of a
normal physiological state. In a pathological state the degree of
chaoticity decreases. A high degree of regularity  is typical of
this condition. Thus, the quantitative estimate of chaoticity in
live systems allows one to define their physiological or
pathological state with a high degree of accuracy. In the
right-hand side of \Ref{f4} we have a statistical invariant, which
reveals the independence of the physical (as well as biophysical,
biochemical and biological) laws in the given live organism  from
the concrete situations as well as the methods of description of
these situations. The invariance, submitted in \Ref{f4}, is
formulated  as the generalization of the experimental data. Among
other physical laws the properties of invariance reflect the most
general and profound properties of the studied systems and
characterize a wide sphere of phenomena. Equation \Ref{f4}
reflects an informational observation. It consists of two
informational measures: the measure of chaoticity and the measure
of pathology (disease).

Let's use the operator of transformation $T (S', S) $ in both
parts of \Ref {f4}. It realizes the transition of the system from
one state $S $ to  other $S' $. By taking into account the
statistical invariance $I (S') = T (S', S) I (S) =I (S) $ in the
right-hand side of \Ref {f4} we get: \bn \Delta
P=P(S')-P(S)=-\Delta C=-\{C(S')-C(S)\}, \Delta P+\Delta C=0.
\label{f5} \en Here the following designations are introduced: $I
(S) $ = $Invariant $, $P (S) =IMP (S) $ is an informational
measure of pathology (disease) for the state $S $, $C (S) =IMC (S)
$ is an informational measure of chaoticity for the state of
patient $S $. Besides in \Ref {f5}  we take into account the rules
of transformation: \be C (S') = T (S', S) C (S), P (S') = T (S',
S) P (S). \label {f6} \ee

Equations \Ref {f4}-\Ref {f6} are rather simple but they make the
quantitative description of the state of a patient possible, both
during the disease and under the medical treatment. Equations \Ref
{f4}-\Ref {f6} have a general character. They are true for many
complex natural and social systems. It is possible to develop the
algorithms of prediction of various demonstrations of chaos in
complex systems of diverse nature on the basis of these equations.
\section{The quantitative factor of quality of a treatment}

One of the major problems of the medical physics consists in the
development of a reliable criterion to estimate the quality of a
medical treatment, a diagnosis and a forecasting of the behaviour
of real live complex systems. As one can see from the previous
section, such a criterion should include the parameter of the
degree of randomness in a live organism. The creation of a
quantitative factor for the quality of a treatment $Q_T $ is based
on the behaviour law of the non-Markovity parameter $
\varepsilon_1 (0) $ in the stochastic dynamics of complex systems.
The greater values of the parameter $ \varepsilon_1 (0) $ are
characteristic of stable physiological states of systems; the
smaller ones are peculiar for pathological states of live systems.
Thus, by the increase or reduction of the non-Markovity parameter
one can judge the physiological state of a live organism with a
high degree of accuracy. Therefore the non-Markovity parameter
allows one to define the deviation of the physiological state of a
system from a normal state.

The factor $Q_T $ defines the efficacy or the quality of the
treatment and is directly connected with the changes in the
quantitative measure of chaoticity in a live organism. We shall
calculate it in a concrete example. Let us consider \textbf {1} as
the patient's state  before therapy, and \textbf {2} as the state
of the patient after certain medical intervention. Then $
\varepsilon_1 (\textbf {1}) $ and $ \varepsilon_1 (\textbf {2}) $
represent quantitative measures of the chaoticity for the
physiological states \textbf {1} and \textbf {2}. The ratio
$\delta$ of these values ($ \delta =\frac {\varepsilon_1 (\textbf
{2})} {\varepsilon_1 (\textbf {1})} $) will define efficacy of the
therapy. Various $j$th processes occur simultaneously in the
therapy. Therefore the total value of $ \delta $ can be defined by
the following way:
\be \delta =\prod _ {j=1} ^ {n} \frac { \varepsilon_1^j (\textbf
{2})} {\varepsilon_1^j (\textbf {1})}, \label {f7} \ee where
$j=1,2... n $ is the number of factors affecting the behaviour of
the non-Markovity parameter. However, the natural logarithm $
\ln\delta $ is  more convenient for use.

Then we have: \bn \delta>1, \ln\delta>0; \nonumber
\\ \delta = 1, \ln\delta = 0; \nonumber
\\ \delta < 1, \ln\delta < 0. \nonumber \en

The three values of $ \delta $ mentioned above correspond to the
three different situations of treatment: effective, inefficient
and destructive treatment. They reflect an increase, preservation
and reduction of the measure of the chaoticity in the therapy.
Thus, one can define $Q_T (\varepsilon) = \ln\delta $ according to
\Ref {f7} as follows: \be Q_T {(\varepsilon)} = \ln \prod _ {j=1}
^ {n} \frac {\varepsilon_1^j (\textbf {2})} { \varepsilon_1^j
(\textbf {1})}. \label {f8} \ee

However, the total factor $Q_T $ is defined both by the
quantitative measures of the chaoticity and by other physiological
and biochemical data. Now we shall consider the transition of the
patient from state \textbf {1} into state \textbf {2}. Then by
analogy, one can  introduce the physiological parameter $k
(\textbf {1}) $, determined for state \textbf {1}, and $k (\textbf
{2}) $ for state  \textbf {2}. In the case of Parkinson's disease
one can introduce the amplitude or the dispersion of the tremor
velocity of some extremities (hand or leg) of the patient as this
parameter. In other cases any medical data, which is considered
for diagnostic purposes, can be used. For greater reliability it
is necessary to use the combination of various parameters $k^j
(\textbf {1}) $ and ($k^j (\textbf {2})$).

The value: \be Q_T =\ln \prod _ {j=1} ^ {n} \frac {\varepsilon_1^j
(\textbf {2})} { \varepsilon_1^j (\textbf {1})} * \left \{\frac
{k^j (\textbf {2})} {k^j (\textbf {1})} \right \}\label {f9} \ee
will be considered as a generalized quantitative factor of quality
of the therapy.

However in real conditions it is necessary to increase or weaken
the magnitude of chaotic, or physiological contributions to \Ref
{f9}. For this purpose we shall take the simple ratio:
\be \ln \prod( a^n b^m ...) =n\ln a+m\ln b +... \nonumber \ee By
analogy, we can reinforce or weaken various contributions
depending on the concrete situation: \be Q_T =\ln \prod _ {j=1} ^
{n} \left (\frac {\varepsilon_1^j (\textbf {2})} { \varepsilon_1^j
(\textbf {1})}\right) ^ {m_j} *\left \{ \frac {k^j (\textbf {2})}
{ k^j (\textbf {1})} \right \} ^ {p_j} .\label {f10} \ee If
incomplete experimental data are available in some situations, one
can assume $p_j=1 $ (attenuation  of the physiological
contribution). A value of $m_j > 1 $ can  mean an amplification of
the chaotic contribution. Otherwise, if we want to weaken the
chaotic contribution, we should take ($m_j=1 $) and if we want to
reinforce the physiological contribution we come towards ($p_j
> 1 $). We have presented  the results of the calculation of the
quantitative factor $Q_T$ below in Sect. 6.

\section{Experimental data}
We have taken the experimental data  from \cite {Beuter1}. They
represent the time records of the tremor velocity of the index
finger of a patients with Parkinson's disease (see, also
http://physionet.org/physiobank/database/). The effect of chronic
high frequency deep brain stimulation (DBS) on the rest tremor was
investigated \cite {Beuter1} in a group of subjects with
Parkinson's disease (PD) (16 subjects). Eight PD subjects with
high amplitude tremor and eight PD subjects with low amplitude
tremor were examined by a clinical neurologist and tested with a
velocity laser to quantify time and frequency domain
characteristics of tremor. The participants received DBS of the
internal globus pallidus (GPi), the subthalamic nucleus (STN) or
the ventrointermediate nucleus of the thalamus (Vim). Tremor was
recorded with a velocity laser under two conditions of DBS
(on-off) and two conditions of medication (L-Dopa on-off).

All the subjects gave informed consent and institutional ethics
procedures were followed. The selected subjects were asked to
refrain from taking their medication at least 12 h before the
beginning of the tests and were not allowed to have more than one
coffee at breakfast on the two testing days. Rest tremor was
recorded on the most affected side with a velocity-transducing
laser \cite{Beuter2, Norm}. This laser is a safe helium-neon
laser. The laser was placed at about 30 cm from the index finger
tip and the laser beam was directed perpendicular to a piece of
reflective tape placed on the finger tip. Positive velocity was
recorded when the subjects extended the finger and negative
velocity when the subjects flexed the finger.

The conditions, counterbalanced across subjects, included the
following:

1. The L-Dopa condition (no stimulation).

2. The DBS condition (stimulation only).

3. The "off" condition (no medication and no stimulation).

4. The "on" condition (on medication and on stimulation).

5. The effect of stopping DBS on tremor (time record of the tremor
after  15, 30, 45, 60 min since switching off of the stimulator).

In Fig. 1  the time records of the velocity of changing tremor of
the index finger of the second patient's hand (man, 52 years old)
under various conditions of  influence on the organism are
submitted as an example. High velocity of tremor is observed: 1)
in a natural condition of the patient (a), 2) 15 (45) minutes
after the stimulator was switched off. Lower tremor speed occurs:
1) when both methods (stimulation, medication) are used, 2) when
each of these methods is used separately, 3) 30 (60) minutes after
the stimulator was switched off. Similar results are presented in
\cite {Beuter1}.

\begin{figure}
\centering
\includegraphics[height=6cm,angle=0]{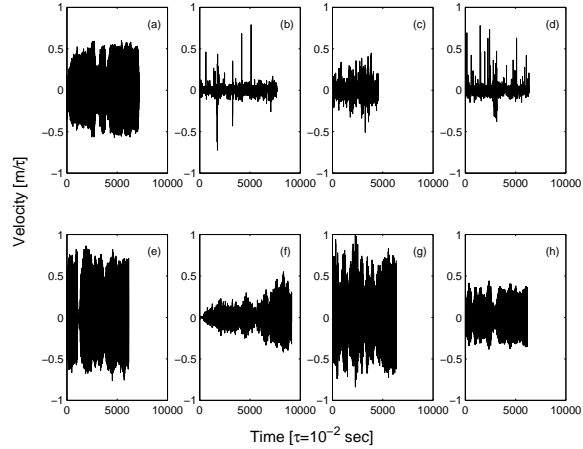}
\caption{The velocity of the change of tremor of the right index
finger of the patient's right hand (the second subject) with
Parkinson's disease under various conditions of the experiment.
\textbf{(a)} deep brain stimulation off, medication off;
\textbf{(b)} the subject was receiving stimulation of the GPi,
medication on; \textbf{(c)} deep brain stimulation off, medication
on; \textbf{(d)} the subject was receiving stimulation of the GPi,
medication off; \textbf{(e)-(h)} the recording of rest tremor in
the right index finger of the subject 15 (30, 45, 60) minutes
after the stimulator was switched off, this subject was off
medication for at least 12 hours} \label{fig:1}
\end{figure}

\section{Results}
In this section the results obtained by processing the
experimental data for one of the patients (subject number 2) are
shown. Similar or related pictures are observed in the
experimental data of other subjects.

\subsection{The non-Markovity parameter as a quantitative
measure of defining chaoticity}

In this subsection the technique to calculate quantitative and
qualitative criteria under various conditions influencing the
state of a patient is given. The basic idea of the approach
consists in defining the quantitative ratio between chaoticity and
regularity of the observed process. It allows one to judge the
physiological (pathological) state of a live system by the degree
of chaoticity or of regularity. The highest degree of chaoticity
in the behaviour of a live system corresponds to a normal
physiological state. Higher degree of regularity or specific
ordering is characteristic of various pathological states of a
live system. In the given work we use the non-Markovity parameter
$ \varepsilon_1 (0) $ as a special quantitative measure defining
chaoticity or regularity of the studied process. The examples
\cite {Yulm1}-\cite {Yulm5}, \cite {Yulm6} which have been
investigated by us earlier serve as a basis for such reasoning. As
one of the examples we shall consider the tremor velocity of the
changing of the subject's index fingers in the case of Parkinson's
disease.

The comparative analysis of the initial time record and the
non-Markovity parameter for all the submitted experimental data
allows one to discover the following regularity. The value of the
non-Markovity parameter  $ \varepsilon_1 (0) $ decreases with the
increase of the tremor velocity  of the patient's fingers
(deterioration of the physiological state) and grows with the
decrease of the tremor velocity (improvement of the state of the
patient). We  shall also consider the power spectra $ \mu_0
(\omega) $ of the initial TCF under various conditions influencing
an organism, the window-time behaviour of the power spectrum  $
\mu_0 (\omega) $ and the non-Markovity parameter $ \varepsilon_1
(\omega) $, the time dependence local averaging relaxation
parameter $ \lambda_1 (t)$ as additional sources of information.

Figure 2 represents the power spectra of the initial TCF for
various experimental conditions. One can observe the powerful peak
in all the figures at the characteristic frequency $ \omega=0.07
f.u. (\omega=2\pi\nu, 1 f.u. = 2\pi/\tau, 1\tau=10^{-2}$ second).
The amplitude values of this peak for $ \mu_i (\omega) $ ($i=1,2,3
$) are given in Table 1. The given peak testifies to a
pathological state of the studied system. A similar picture is
observed in patients with myocardial infarction \cite {Yulm2}. The
comparison of these values reflects the amplitude of the tremor
velocity at the initial record of time.

\begin{figure}
\centering
\includegraphics[height=6cm,angle=0]{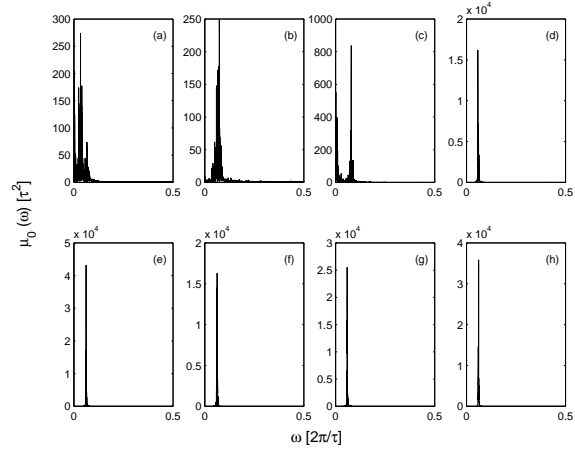}
\caption{The power spectrum $ \mu_0 (\omega) $ of the initial TCF
for the velocity of changing of tremor of subject number 2 under
various conditions that influence an organism. \textbf{(a)} deep
brain stimulation on, medication on; \textbf{(b)} deep brain
stimulation on, medication off; \textbf{(c)} deep brain
stimulation off, medication on; \textbf{(d)} deep brain
stimulation off, medication off; \textbf{(e)-(h)} the power
spectrum $ \mu_0 (\omega) $ of the initial TCF  for the recording
of rest tremor in the right index finger of  the subject  15 (30,
45, 60) minutes after the stimulator was switched off, medication
off. At the frequency $ \omega = 0.07 f. u., 1 f. u.=100 Hz $ (the
characteristic frequency), a peak is found. The presence and
amplitude of this peak are determined by the state of the patient}
\end{figure}

\bigskip
\footnotesize \begin {center} \textbf{Table 1.} The value
$\mu_0(\omega)$ for the initial TCF and $\mu_i(\omega)$ ($ i=1, 2,
3$) for the memory functions of junior orders at the frequency
$\omega=0.07 f.u.$ 1 - Deep brain stimulation, 2 - Medication
(subject number 2). For example, OFF OFF - no DBS and no
medication. \label{tab:1} \tiny
\begin{tabular} {p{1.0cm}p{1.0cm}p{1.0cm}p{1.0cm}p{1.2cm}p{1.2cm}p{1.2cm}p{1.2cm}p{1.2cm}}
\\ \hline
\ \  & ON ON & ON OFF & OFF ON & OFF OFF & 15 OFF& 30 OFF& 45 OFF& 60 OFF \\
\hline
$\mu_0$ & 75 & 250 & 812 & $1.71*10^4$ & $4.34*10^4$ & $1.53*10^4$ & $2.51*10^4$ & $3.68*10^4$ \\
$\mu_1$ & 19  & 52 & $1.28*10^3$ & $1.17*10^4$  & $3.21*10^4$ & $1.32*10^4$ & $1.82*10^4$ & $2.8*10^4$ \\
$\mu_2$ & 42 & 60 & 113 & 71& 300 & 62& 137& 224 \\
$\mu_3$ & 37 & 54 & 141 & 73& 147 & 74& 152& 186 \\
\hline
\end{tabular}\end {center}
\bigskip
\normalsize \

In Table 1 the second patient's amplitude values $ \mu_0 (\omega)
$ for the initial TCF and the memory functions of junior orders $
\mu_i (\omega) $ ($ i=1, 2, 3 $) at the frequency $ \omega=0.07
f.u. $ are submitted. The terms of the first row define the
conditions under which the experiment is carried out. Under all
conditions a power peak at the frequency $ \omega=0.07 f.u. $ can
be observed. The amplitude values of the given peak (in particular
in the power spectrum $ \mu_0 (\omega) $) reflect the amplitude of
the tremor velocity. For example, the least amplitude 75 $ \tau^2
$ corresponds to the condition (ON, ON; or: deep brain stimulation
on, medication on). The highest amplitude $4.34*10^4 \tau^2 $
corresponds to the greatest tremor speed (see Figs. 1e, 2e). Thus,
the given parameter can be used to estimate the physiological
state of a patient. A similar picture is observed in all other
patients.

In Fig. 3 the initial time record (the normal state of the
subject; OFF, OFF) and the window-time behaviour of the power
spectrum of the TCF (the technique of the analysis of the given
behaviour is considered in Ref. \cite {Yulm6}) are submitted. In
these figures regions 1, 2, 3, which correspond to the least
values of the tremor velocity are shown. The minimal amplitude of
the peaks of the power spectrum $ \mu_0 (\omega)$ corresponds to
the regions with the least tremor velocity.

\begin{figure}
\centering
\includegraphics[height=6cm,angle=0]{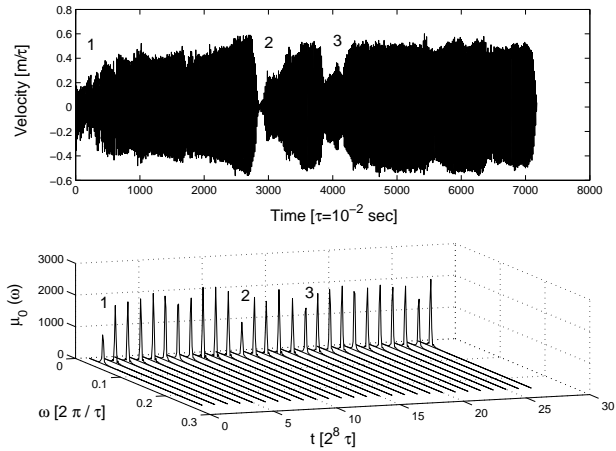}
\caption{The initial time series and the window-time behaviour of
the power spectrum $ \mu_0 (\omega) $ of the TCF. Two figures are
submitted to illustrate the case of subject number 2: no
stimulation of the brain and no medication. The change of regimes
in the initial time series is reflected in the decrease of the
tremor velocity (regions 1, 2 and 3) and becomes visible as a
sharp reduction of the power of spectrum $ \mu_0 (\omega) $ (see,
the $1th, 12th, 17 th $ windows for more detail)}
\end{figure}
\begin{figure}
\centering
\includegraphics[height=6cm,angle=0]{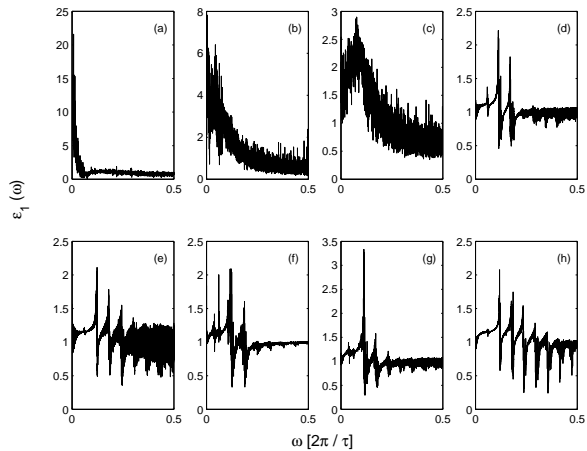}
\caption{The first point of the non-Markovity parameter  $
\varepsilon_1 (\omega) $ for the second  subject under various
experimental conditions: \textbf{(a)} deep brain stimulation off,
medication on; \textbf{(b)} deep brain stimulation on, medication
on; \textbf{(c)} deep brain stimulation on, medication off;
\textbf{(d)} deep brain stimulation off, medication off;
\textbf{(e)-(h)} the recording of rest tremor in the right index
finger of the subject 15 (30, 45, 60) minutes after the stimulator
was switched off, medication off. The non-Markovity parameter at
zero frequency $\varepsilon_1(0)$ plays a special role. These
values (6.02 in the first case and 1.0043 in the last one) define
chaoticity or the regularity of the studied states. The amplitudes
of these values also characterize the state of the subject}
\end{figure}

In Fig. 4 the frequency dependence of the first point of the
non-Markovity parameter  $ \varepsilon_1 (\omega) $ is submitted
for the second subject  under various experimental conditions. The
value of the parameter  $ \varepsilon_1 (0) $ at zero frequency is
of special importance for our study of manifestations of
chaoticity. It is possible to judge the change of the state of a
subject by the increase (or by the decrease) of this value. The
comparative analysis of the initial time records allows one to
come to the similar conclusions. In Figs. 4d-h a well-defined
frequency structure of the non-Markovity parameter can be seen.
The structure is completely suppressed and disappears only when
during the treatment. The characteristic frequency of the
fluctuations corresponds approximately to $ \omega=0.06 f.u. $
These multiple peaks are the most appreciable at low frequencies.
At higher frequencies these fluctuations are smoothed out. As can
be seen in these figures, the 2nd subject has a strong peak which
remains stable over time. As our data show, the comb-like
structure with multiple frequencies can be observed in all
patients with high tremor velocity. In a group of patients with
low tremor velocity it disappears, and  a wider spectrum that
presents some fluctuations over time is observed. The present
structure testifies to the presence of characteristic frequency of
fluctuations of tremor of human extremities.

In Table 2 the dispersion interval of the values and the average
value $ \varepsilon_1 (0) $ for the whole group of subjects (16
subjects) are submitted. Let us consider 2 conditions: OFF, OFF
and OFF, ON. In the first case the dispersion interval and the
average value $ \varepsilon_1 (0) $ are minimal. It means the
presence of a high degree of regularity of the physiological state
of the patient. The degree of regularity is appreciably reduced
when applying any method of treatment. Here the degree of
chaoticity grows. The maximal degree of chaoticity corresponds to
the condition OFF, ON (medication is used only). The difference in
$ \varepsilon_1 (0) _ {av.val} $ with medication and without it
(OFF, OFF) is 3.8 times (!). On the basis of the comparative
analysis of the given parameters the best method of treatment for
each individual case can be found. It is necessary to note, that
the given reasoning is true only for the study of the chaotic
component of the quantitative factor of the quality of treatment
$Q_T $. The most trustworthy information about the quality of
treatment can be given by the full quantitative factor $Q_T $
which takes into account of other diagnostic factors.

\bigskip \footnotesize \begin{center} \textbf{Table 2.} The dispersion interval
$ \varepsilon_1 (0) _ {int} $ of the values and the average value
$ \varepsilon_1 (0) _ {av.val} $ of the first point of the
non-Markovity parameter  under various experimental conditions for
the group of 16 subjects. 1 - Deep brain stimulation, 2 -
Medication.

\tiny
\begin{tabular}{p{1.5cm}p{1.2cm}p{1.2cm}p{1.0cm}p{1.0cm}p{1.0cm}p{1.0cm}p{1.0cm}p{1.0cm}}
\\ \hline
\ \  & OFF OFF & ON OFF & OFF ON & ON ON & 15 OFF& 30 OFF& 45 OFF& 60 OFF \\
\hline
$\varepsilon_1(0)_{int}$ & 1 - 1.8 & 2 - 18 & 2 - 22 & 1.5 - 8 & 1.5 - 3 & 1.8 - 5 & 1.7 - 4.5 & 2 - 6 \\
$\varepsilon_1(0)_{av.val}$ & 1.41  & 4.14 & 5.31 & 3.17 & 2.43 & 2.92 & 2.76 & 2.93\\
\hline
\end{tabular}
\end{center}
\bigskip
\normalsize

The results of the calculation of the quantitative factor $Q_T$
are shown in Table 3. The data are submitted for a single patient
and for the whole group. Here $Q_T(\varepsilon)$ is the chaotic
contribution to the quantitative factor (see \Ref{f8}). $Q_T$ is
the total quantitative factor (see \Ref{f10}), where
$\varepsilon^{(1)}(\textbf{1})$ and
$\varepsilon^{(1)}(\textbf{2})$ are the chaotic contributions for
the tremor amplitudes $k^{(1)}(\textbf{1}), k^{(1)}(\textbf{2})$;
$\varepsilon^{(2)}(\textbf{1})$ and
$\varepsilon^{(2)}(\textbf{2})$ are the dispersions of the tremor
amplitudes $k^{(2)}(\textbf{1}), k^{(2)}(\textbf{2})$
(physiological contributions). The full factor $Q_T $ provides
detailed information about the quality of the treatment. The
present factor includes both the chaotic component $Q_T
(\varepsilon) $, and the physiological contribution $Q_T (k) $.
The calculation $Q_T (k) $ is described in Sect. 4. One can define
the quality of a treatment by means of $Q_T $. The positive value
of the given factor defines an effective treatment. For a separate
patient and for the whole group, $Q_T $ reaches its maximal value
under the condition  ON, ON. The total quantitative factor is
supplemented by a diagnostic (physiological) component. It allows
one to take into account those features of the system which the
chaotic component does not contain. For the second patient under
condition 15 OFF (see Table 3) the factor $Q_T $ has a negative
value. It testifies to the negative influence of the given
treatment on the organism of the patient. The best treatment is
thus the combination of the two medical methods: electromagnetic
stimulation and medication.

\bigskip \footnotesize \begin{center} \textbf{Table 3.} The quantitative factor
$Q_T(\varepsilon)$ and the total quantitative factor $Q_T$ for the
second patient and for the whole group (16 subjects). 1 - Deep
brain stimulation, 2 - Medication. $m_j=1, p_j=1.$

\tiny
\begin{tabular}{p{1.0cm}p{1.2cm}p{1.2cm}p{1.2cm}p{1.5cm}p{1.0cm}p{1.0cm}p{1.0cm}p{1.0cm}}

\\ \hline  &&&& The 2 patient \\
\ \  & OFF OFF & ON OFF & OFF ON & ON ON & 15 OFF& 30 OFF& 45 OFF& 60 OFF \\
\hline
$Q_T(\varepsilon)$ & & 0.758& 2.556& 1.756& 0.291& 0.438& 0.041& 0.017\\
$Q_T$ & & 1.763& 2.013& 2.654 & -0.013 & 0.883 & -0.004& 0.856\\
\hline &&&& The all group \\
$Q_T(\varepsilon)$ & & 1.077& 1.326&0.810& 0.544& 0.728& 0.671& 0.731\\
$Q_T$ & & 3.661& 2.883& 4.071& 1.47 & 1. 734& 1.624& 1.742 \\
\hline
\end{tabular}
\end{center}
\bigskip
\normalsize

Figure 5 reflects the behaviour of the parameter $ \varepsilon_1
(0) $ for four different patients. The points lying above the
horizontal line testify to an improvement of the state of the
subject and the efficacy of the treatment. The points, lying below
the horizontal line testify to a deterioration of the state of the
subject and the inefficiency of the applied method. For example,
Fig. 5b corresponds to the sevenfold change of the quantitative
measure of chaoticity for the 9th patient.  In case of the 8th
patient  (see Fig. 5c) no influence could change the measure of
the chaoticity. Therefore there was practically no change in the
state of the subject either. In some cases (see Figs. 5b, 5d) the
DBS or the medication reduces the measure of chaoticity which
testifies to a deterioration of the state of the subject. This
approach allows one to define the most effective (or inefficient)
treatment in each individual case.

\begin{figure}
\centering
\includegraphics[height=6cm,angle=0]{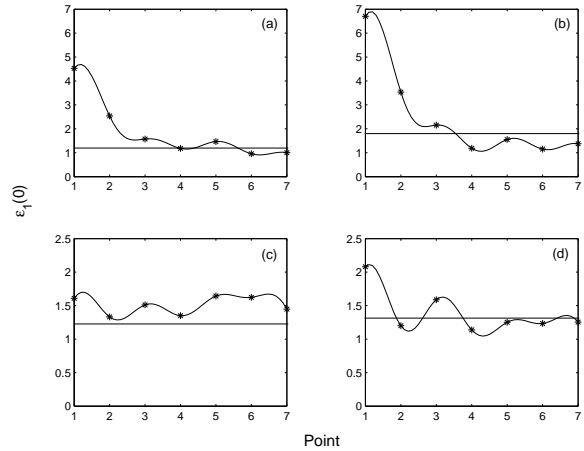}
\caption{The behavior of the parameter $ \varepsilon_1 (0) $ for
four various patients: \textbf{(a)} the second subject, high
amplitude tremor, \textbf{(b)} the 9th subject, low amplitude
tremor (the stimulation of the GPi); \textbf{(c)} the 8th subject,
high amplitude tremor, \textbf{(d)} the 15th subject, low
amplitude tremor (the stimulation of the STN). The value of $
\varepsilon_1 (0) $ for seven consecutive conditions of the
experiment: 1 - both methods are used; 2 - treatment by medication
is applied only; 3 - the DBS  is used only; 4 (5, 6, 7) - value of
the parameter 15 (30, 45, 60) minutes after the stimulator was
switched off; the horizontal line corresponds to the value of the
parameter when no method is used. This representation allows one
to define the most effective treatment for each patient}
\end{figure}
\subsection{The definition of a predictor of sudden changes of the tremor velocity}

In this subsection  the window-time behaviour of the non-Markovity
parameter  $ \varepsilon_1 (\omega) $ for a certain case (the
second patient, two methods of medical treatment were used) and
the procedure of local averaging of the relaxation parameters are
considered. These procedures allow one to determine specific
predictors of the change of regimes in the initial time records.

The idea of the first procedure is, that  the optimum length of
the time window ($2^8 = 256 $ points) is   found first. In the
studied dependence (in our case  the frequency dependence of the
first point of the non-Markovity parameter) the first window is
cut out. Then the second window is cut out  (from point 257 points
to point 512), etc. This construction allows one to find the local
time behaviour of the non-Markovity parameter. At the critical
moments when the tremor velocity increases the value of the
non-Markovity parameter comes nearer to a unit value. One can
observe that the value of the non-Markovity parameter starts to
decrease 2-2.5 sec before the increase of the tremor velocity (see
Fig. 6).
\begin{figure}
\centering
\includegraphics[height=6cm,angle=0]{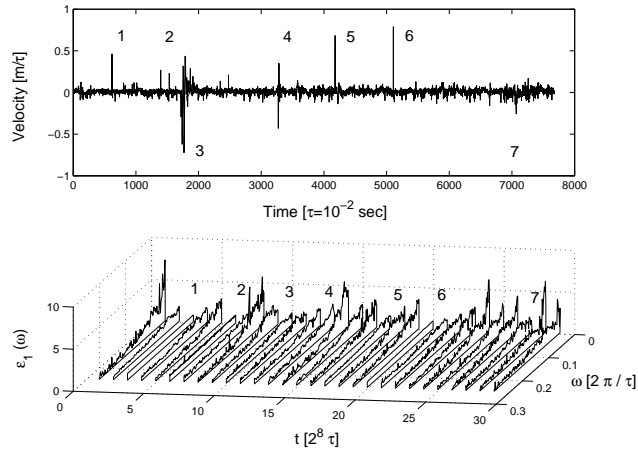}
\caption{The initial signal - the change of tremor velocity when
the second patient is treated by two methods - and the window-time
behaviour of the first point of the non-Markovity parameter $
\varepsilon_1 (\omega) $. At the time of a sharp change of the
mode (sharp increase of the tremor velocity) in the behaviour of
the initial time series (regions 1-7) a gradual decrease of the
non-Markovity parameter down to a unit value (the $3rd, 6th, 10th,
14th, 17th, 20th, 27th $ windows) is observed. The decrease of the
non-Markovity parameter begins 2-2,5s earlier that the tremor
acceleration on an initial series}
\end{figure}

\begin{figure}
\centering
\includegraphics[height=6cm,angle=0]{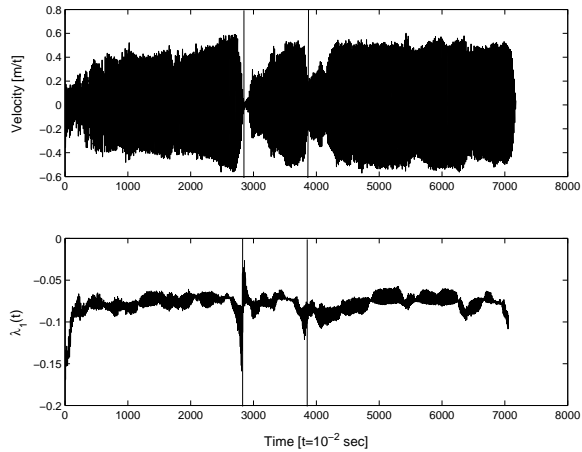}
\caption{The change of the tremor velocity for the second patient
(stimulation of the brain and medication are not used) and the
time dependence of the local relaxation parameter $ \lambda_1 (t)
$. The localization procedure allows one to find sudden changes of
relaxation regimes of the systems under consideration. The largest
amplitude values of the local relaxation parameter are in the
region of the lowest tremor velocity. The change of the time
behaviour of the parameter $ \lambda_1 (t) $ begins 2-3s earlier
than the sharp change of the regimes in the initial time series
appears}
\end{figure}

The idea behind the second procedure is the following: one can
consider the initial data set  and take an \emph{N}-long sampling.
We can calculate kinetic and relaxation parameters for the given
sampling. Then we can carry out a "step-by-step shift to the
right". Then we calculate kinetic and relaxation parameters. After
that we execute one more "step-by-step shift to the right" and
continue the procedure up to the end of the time series. Thus, the
local averaged parameters have a high sensitivity to the effects
of intermittency and non-stationarity. Any non-regularity in the
initial time series is instantly reflected in the behaviour of the
local average parameters. The optimal length of the sample is 120
points. In Figs. 7, 8 the initial time record and the time
dependence of the local relaxation parameter $ \lambda_1 (t) $ are
shown in two cases. The change in the time behaviour of the
parameter $ \lambda_1 (t) $ begins 2-3s prior to the change of the
regimes of the time record of the tremor velocity. The increase of
speed of the local relaxation parameter $ \lambda_1 (t) $
testifies to a decrease of tremor velocity.

\begin{figure}
\centering
\includegraphics[height=6cm,angle=0]{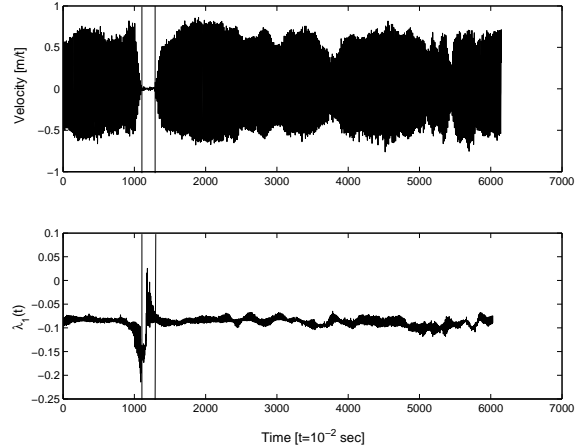}
\caption{The change of the tremor velocity for  the second patient
(15 minutes after the stimulator was switched off, medication off)
and the time dependence of the local relaxation parameter $
\lambda_1 (t) $. The site characterizing the minimal tremor
velocity is allocated. The increase and decrease of the local
relaxation parameter occurs 2.5s before the decrease or increase
of the tremor velocity. The similar behaviour of the parameter $
\lambda_1 (t) $ can be explained by its high sensitivity to the
presence of nonstationarity of the initial signal}
\end{figure}

\section{Conclusions}

In this chapter we have proposed a new concept for the study of
manifestations of chaoticity. It is based on the application of
the statistical non-Markovity parameter  and its spectrum as an
informational measure of chaoticity. This approach allows one to
define the difference between a healthy person and a patient by
means of the numerical value of the non-Markovity parameter. This
observation gives a reliable tool for the strict quantitative
estimates that are necessary for the diagnosis and the
quantification of the treatment of patients. As an example we have
considered the changes of various dynamic conditions of patients
with Parkinson's disease. The quantitative and qualitative
criteria used by us for the definition of chaoticity and
regularity of investigated processes in live systems reveal new
informational opportunities of the statistical theory of discrete
non-Markov random processes. The new concept allows one to
estimate quantitatively the efficacy and the quality of the
treatment of different patients with Parkinson's disease. It
allows one to investigate various dynamic states of complex
systems in real time.

The statistical non-Markovity parameter $ \varepsilon_1 (0) $ can
serve as a reliable quantitative informational measure of
chaoticity. It allows one to use $ \varepsilon_1 (0) $ for the
study of the behaviour of different chaotic systems. In the case
of Parkinson's disease the change of the parameter defines the
change of a quantitative measure of chaosity or regularity of a
physiological system. The increase of chaoticity reflects the
decrease of the quantitative measure of pathology and the
improvement of the state of the patient. The increase of the
regularity defines high degree of manifestation of pathological
states of live systems. The combined power spectra of the initial
TCF $ \mu_0 (\omega) $, the three memory  functions of junior
orders and the frequency dependence of the  non-Markovity
parameter compose an informational measure which defines the
degree of pathological changes in a human organism.

The new procedures (the window-time procedure and the local
averaging procedure) give  evident predictors of the change of the
initial time signal. The window-time behaviour of the
non-Markovity parameter $ \varepsilon_1 (\omega) $ reflects the
increase of the tremor velocity 2-2.5s earlier. It happens when
the non-Markovity parameter approaches a unit value. The procedure
of local averaging of the relaxation parameter $ \lambda_1 (t) $
reflects the relaxation changes of physiological processes in a
live system. The behaviour of the local parameter $ \lambda_1 (t)
$ reacts to a sudden change of relaxation regimes in the initial
time record 2-3s earlier. These predictors allow to lower the
probability of ineffective use of different methods of treatment.

In the course of the study we have come to the following
conclusions:
\\ - The application of medication for the given group of patients
proved to be the most efficient way to treat patients with
Parkinson's disease. Used separately stimulation is less effective
than the the use of medication.
\\ - The combination of different methods (medication plus electromagnetic
stimulator) is less effective than the application of medication
or of stimulation. In some cases the combination of medication and
stimulation exerts a negative influence on the state of the
subject.
\\ - After the stimulator is switched off its
aftereffect has an oscillatory character with a  low
characteristic frequency corresponding to a period of 30 min.
\\- The efficacy of various medical
procedures and the quality of a treatment can be estimated
quantitatively for each subject separately with utmost precision.
\\- \textbf{However, if we take  both chaotic and physiological components
into account, the general estimation of the quality of treatment
will be more universal. The combination of two methods (DBS and
medication, $Q_T=4.071$) produces the most effective result in
comparison with the effect of DBS (3.661) or of medication (2.883)
given separately. This is connected with additional aspect of the
estimation of the quality of treatment due to the study both of
both chaotic and diagnostic components of a live system. This
conclusion corresponds to the  results of \cite {Beuter1}.}

In conclusion we would like to state that our study gives a unique
opportunity for the exact quantitative description of the states
of patients with Parkinson's disease at various stages of the
disease as well as of the treatment and the recovery of the
patient. On the whole, the proposed concept of manifestations of
chaoticity opens up great opportunities for the alternative
analysis, diagnosis and forecasting of the chaotic behavior of
real complex system of live nature.

\section{Acknowledgements}
This work  supported by the RHSF (Grant No. 03-06-00218a), RFBR
(Grant No. 02-02-16146) and CCBR of Ministry of Education RF
(Grant No. E 02-3.1-538). The authors acknowledge Prof. Anne
Beuter for stimulating criticism and valuable discussion and Dr.
L.O. Svirina for technical assistance.

\end{document}